\documentclass{pasj01}
\Received{$\langle$reception date$\rangle$}
\Accepted{$\langle$acception date$\rangle$}
\Published{$\langle$publication date$\rangle$}

\begin{document}

\title{Mass constraint for a planet in a protoplanetary disk from the gap width}
\author{Kazuhiro D. KANAGAWA\altaffilmark{1,2}, 
Takayuki MUTO\altaffilmark{3}, 
Hidekazu TANAKA\altaffilmark{1}, 
Takayuki TANIGAWA\altaffilmark{4}, 
Taku TAKEUCHI\altaffilmark{5},
Takashi TSUKAGOSHI\altaffilmark{6},
Munetake MOMOSE\altaffilmark{6}}%
\altaffiltext{1}{Institute of Low Temperature Science, Hokkaido
University, Sapporo 060-0819, Japan} 
\altaffiltext{2}{Institute of Physics and CASA$^{\ast}$, Faculty of Mathematics and Physics, University of Szczecin, Wielkopolska 15, 70-451 Szczecin, Poland} 
\altaffiltext{3}{Division of Liberal Arts, Kogakuin University, 1-24-2,
Nishi-Shinjuku, Shinjuku-ku, Tokyo, 163-8677, Japan} 
\altaffiltext{4}{School of Medicine, University of Occupational and
Environmental Health, Yahatanishi-ku, Kitakyushu, Fukuoka 807-8555,
Japan} 
\altaffiltext{5}{Department of Earth and Planetary Sciences, Tokyo
Institute of Technology, Meguro-ku, Tokyo 152-8551, Japan} 
\altaffiltext{6}{College of Science, Ibaraki University,  
2-1-1, Bunkyo, Mito, Ibaraki 310-851, Japan} 
\email{kazuhiro.kanagawa@usz.edu.pl}


\maketitle

\begin{abstract}
A giant planet creates a gap in a protoplanetary disk, which might explain the observed gaps in protoplanetary disks.
The width and depth of the gaps depend on the planet mass and disk properties.
We have performed two--dimensional hydrodynamic simulations for various planet masses, disk aspect ratios and viscosities, to obtain an empirical formula for the gap width.
The gap width is proportional to the square root of the planet mass, $-3/4$ power of the disk aspect ratio and $-1/4$ power of the viscosity.
This empirical formula enables us to estimate the mass of a planet embedded in the disk from the width of an observed gap.
We have applied the empirical formula for the gap width to the disk around HL~Tau, assuming that each gap observed by ALMA observations is produced by planets, and discussed the planet masses within the gaps.
The estimate of planet masses from the gap widths is less affected by the observational resolution and dust filtration than that from the gap depth.
\end{abstract}

\section{Introduction}
Recent observations of protoplanetary disks have shown that disks with non--axisymmetric structures (e.g., \cite{Casassus2013,Fukagawa2013,vanderMarel2013,Perez2014}) and/or  gap structures (e.g.,\cite{Osorio2014,ALMA_HLTau2015}).
One possible origin of these structures are the dynamic interaction between the disk and embedded planets \citep{Lin_Papaloizou1979,Goldreich_Tremaine1980,Lin_Papaloizou1993}.
A large planet embedded in a disk produces a gap around its orbit.
The planet mass and the disk properties are reflected in the gap width and depth.
It is important to construct a model of a gap that can predict a planet mass.

Recent studies on the gap formation (e.g.,\cite{Duffell_MacFadyen2013,Kanagawa2015a, Kanagawa2015b}, hereafter Paper~I) have revealed that the gap depth is related with the planetary mass, the disk aspect ratio (temperature), and the viscosity, as
\begin{equation}
	\frac{\Sigma_{\min}}{\Sigma_0}=\frac{1}{1+0.04K},
	\label{eq:smin}
\end{equation}
where $\Sigma_{\min}$ and $\Sigma_0$ are the surface densities at the bottom of the gap and the edge, respectively.
The dimensionless parameter $K$ is defined by
\begin{equation}
	K\equiv \left( \frac{M_p}{M_{\ast}} \right)^{2}\left( \frac{h_p}{R_p} \right)^{-5} \alpha^{-1},
	\label{eq:k}
\end{equation}
where, $M_{p}$, $M_{\ast}$, $R_p$, $h_p$ $\alpha$ are the masses of the planet and the central star, the orbital radius of the planet, the scale height at $R_p$, and the viscous parameter with the prescription by \cite{Shakura_Sunyaev1973}, respectively.
The gap depth given by Equation~(\ref{eq:smin}) agrees well with the results of the hydrodynamic simulations \citep{Varniere_Quillen_Frank2004,Duffell_MacFadyen2013,Fung_Shi_Chiang2014}.

As seen from Equation~(\ref{eq:smin}), the gap depth is determined by the dimensionless parameter $K$, which is a function of $M_p$, $h_p$ and $\alpha$.
Hence, the planet mass can be estimated from the depth of the observed gap if the disk aspect ratio and the viscosity are given.
Paper~I applied Equation~(\ref{eq:smin}) to a gap of the HL~Tau disk observed by the ALMA Long Baseline Campaign \citep{ALMA_HLTau2015} and estimated that the lower--limit of the mass of the planet within the gap at $30$AU is $0.3M_J$ if this gap is originated from the disk--planet interaction.
It is, however, very difficult to estimate the mass of a planet in a deep gap because the emission at the bottom of the gap should be measured with a reasonable signal--to--noise ratio.
In contrast, the gap width can be more easily measured than the gap depth.

It is known that the width of the gap induced by a planet increases with the planet mass \citep{Takeuchi_Miyama_Lin1996, Varniere_Quillen_Frank2004,Duffell_MacFadyen2013,Duffell2015}.
However, a quantitative relationship between the gap width and the planet mass is not clear.
\cite{Varniere_Quillen_Frank2004} reported that if $(M_p/M_{\ast})^{2} (h_p/R_p)^{-2} \alpha^{-1} \gtrsim 0.3$, the gap edges are between the locations of the $m=2$ and $1$ outer Lindblad resonances.
If Keplerian rotation is assumed, the distances between the planet and the $m=2$ and $1$ outer Lindblad resonances are $0.31R_p$ and $0.59R_p$ , respectively \citep{Goldreich_Tremaine1980}.
On the other hand, the hydrodynamic simulations performed by \cite{Duffell_MacFadyen2013} shows that narrower gaps are created.
The half width of the gap given by \cite{Duffell_MacFadyen2013} is below $0.23R_p$ even if $(M_p/M_{\ast})^{2} (h_p/R_p)^{-2} \alpha^{-1} > 0.3$ (see figure~{6} of their paper).
Further investigation have been required in order to constrain the planet mass from the width of the observed gaps.

In this Paper, we derive an empirical relationship between the gap width and the planet mass, performing $26$ runs of two--dimensional hydrodynamic simulation.
In Section~\ref{sec:method}, we describe the numerical method.
In Section~\ref{sec:results}, we show our results and the empirical formula for the gap width.
We apply the formula to estimate the masses of the planets in the observed gaps of the HL~Tau disk in Section~\ref{sec:applications}.
Section~\ref{sec:summary} is for summary.

\section{Numerical method} \label{sec:method}
We study the shape of the gap induced by a planet embedded in a protoplanetary disk using the two--dimensional hydrodynamic code FARGO \citep{Masset2000}, which is widely used to study the disk--planet interaction (e.g.,\cite{Crida_Morbidelli2007,Baruteau_Meru_Paardekooper2011,Zhu2011}).
The computational domain ranges from $R/R_p=0.4$ to $4.0$ with $1024 \times 2048$ radial and azimuthal zones.
The disk scale height, $h$, is resolved by $22$ (radial) and $16$ (azimuthal) zones in the vicinity of the planet.
For simplicity, we neglect the gas accretion onto the planet and the planets rotates on the fixed orbit with $R=R_p$.
We adopt a constant kinematic viscosity coefficient $\nu$, which is $\nu=\alpha c_p h_p$, \citep{Shakura_Sunyaev1973}, where $c_p$ is sound speed at $R=R_p$.
The disk aspect ratio $h/R$ is also set to be a constant throughout the disk.
We adopted a smoothing length for a gravitational potential of a planet as $0.6 h_p$.
We have checked that the choice of the smoothing length does not significantly influence the gap width.

We perform $26$ runs of the hydrodynamic simulation for various planetary masses ($0.1 M_J$~--~$2 M_J$, if $M_{\ast}=1M_{\odot}$), disk aspect ratios ($1/30$~--~$1/15$) and the parameter $\alpha$ of the viscosity ($10^{-2}$~--~$10^{-4}$), which are listed in Table~\ref{tab:models}.
In this work, we follow $10^{4}$ -- $10^{5}$ orbits at the planet's location to reach the steady state.
In the cases with $\alpha=10^{-4}$, a very long time, i.e., $\sim 10^{5}$ planetary orbits, is required to obtain the steady gap width.
Such long calculations are necessary because of the slow viscous evolution in the less viscous disk.

Initially, the surface density is constant ($\Sigma(R)=\Sigma_0$) in the whole region.
The initial angular velocity is given as $\Omega_{\rm K} \sqrt{1-\eta}$, where $\Omega_{\rm K}$ is the Keplerian angular velocity and $\eta = (1/2)(h/R)^2 d\ln P/d\ln R$.
The radial drift velocity is given by $v_R = -3\nu/(2R)$.
The planet mass smoothly builds up from zero to the final value by using the ramp function defined by $\sin^2 \left[ \pi t/ \left( 64 {\rm P}_{\rm orbit} \right) \right]$.

At the inner and outer boundaries($R/R_p=0.4$ and $4.0$), we keep the initial condition described above.
In addition, we introduce wave-killing zones near the boundaries ($0.4<R/R_p<0.5$ and $3.2<R/R_p<4.0$) to avoid artificial wave reflection at the boundaries \citep{Val-Borro_etal2006}.

\section{Result} \label{sec:results}
\subsection{Empirical formula for the gap width}
\begin{figure*}
	\begin{center}
		\resizebox{0.98\textwidth}{!}{\includegraphics{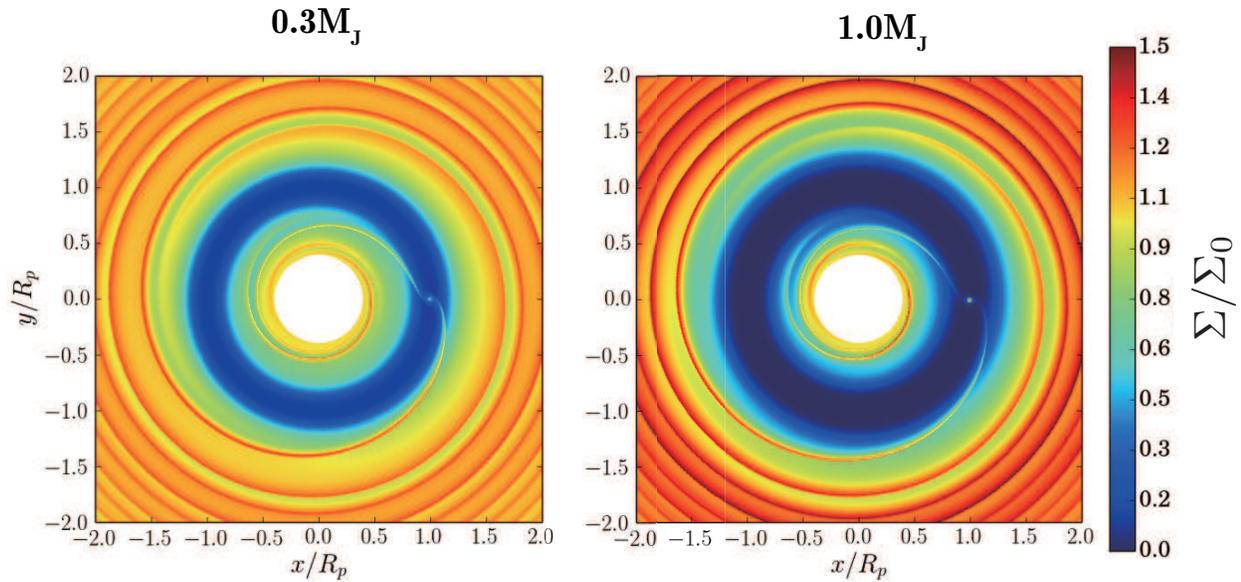}}
	\end{center}
		\caption{
		The surface density distributions at $10^{4}$ planetary orbits obtained by two--dimensional hydrodynamic simulations for $M_p=0.3M_J$ ({\it left}) and $M_p=1.0M_J$ ({\it right}).
		Other parameters are set to be $h_p/R_p=1/20$, $\alpha=10^{-3}$ and $M_{\ast}=1M_{\odot}$.
		\label{fig:densmap}
		}
\end{figure*}
\begin{figure}
	\begin{center}
		\resizebox{0.49\textwidth}{!}{\includegraphics{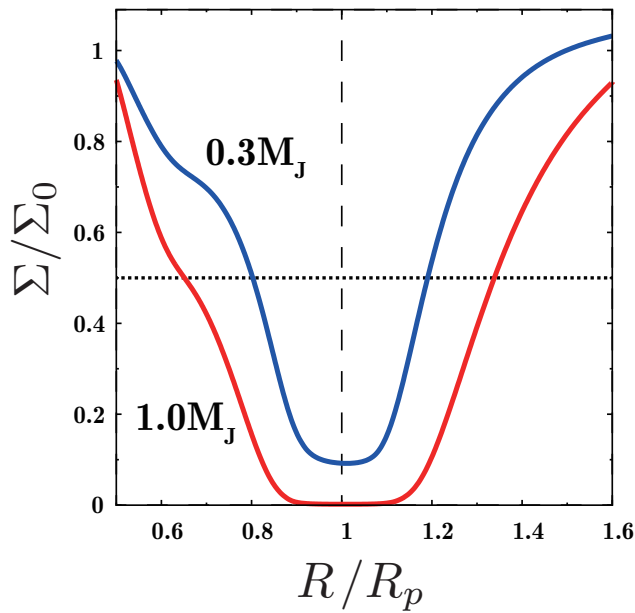}}
	\end{center}
		\caption{
		The radial distributions of the azimuthally averaged surface density for runs with $M_p=0.3M_J$ (blue) and $1.0M_J$ (red) presented in Figure~\ref{fig:densmap}.
		The horizontal dotted line indicates the level of $\Sigma_0/2$.
		\label{fig:avgdens}
		}
\end{figure}
Figure~\ref{fig:densmap} shows two--dimensional distributions of the surface density at $t=10^{4}$ planetary orbits in runs with $M_p=0.3M_J$ and $1.0M_J$.
To measure the gap width, we take an azimuthal average of the surface density (Figure~\ref{fig:avgdens}).
We define the gap region by the radial extent where the azimuthally averaged surface density is smaller than the half of the initial surface density.
The gap width $\Delta_{\rm gap}$ is given by $R_{\rm out}-R_{\rm in}$.
Then we obtain radii $R_{\rm in}$, $R_{\rm out}$ of the inner and outer edges of the gap region.
, where $R_{\rm out}$ and $R_{\rm in}$ are the radii of the outer and inner edges of the gap region. 
Note that we can make a reasonable guess of the gap width from only the snapshot of the simulations (or observations) since the surface density approaches $\Sigma_0$ outside the gap region.
The gap width in the case of $1.0M_J$ ($\Delta_{\rm gap} = 0.69 R_p$) is $\sim 80\%$ larger than that in the case of $0.3M_J$ ($\Delta_{\rm gap}=0.39R_p$).

\begin{figure}
	\begin{center}
		\resizebox{0.49\textwidth}{!}{\includegraphics{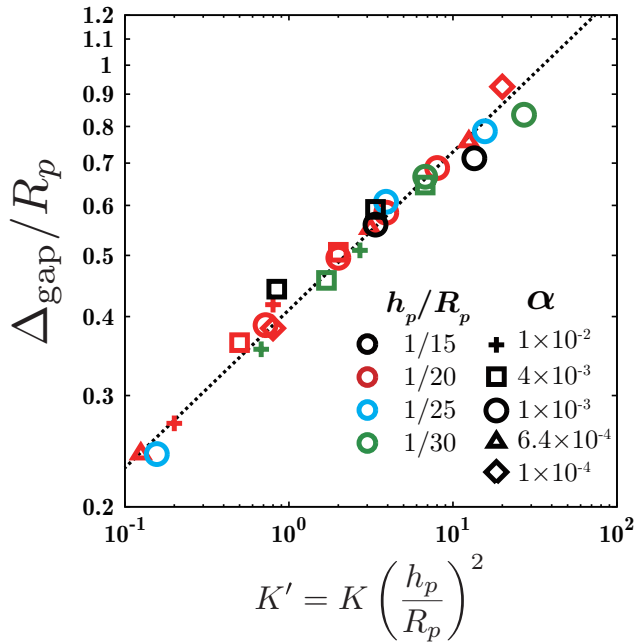}}
	\end{center}
		\caption{
		The widths of the gaps, $\Delta_{\rm gap}=(R_{\rm out}-R_{\rm in})$, against the dimensionless parameter $K'$.
		The dotted line is the empirical formula for the gap width given by Equation~(\ref{eq:gapwidth}).
		The color of symbol denotes the disk aspect ratio (black, red, blue and green indicate $h_p/R_p=1/15$, $1/20$, $1/25$ and $1/30$, respectively).
		The symbol denote the viscosity (cross, square, circle, triangle and diamond indicate $\alpha=10^{-2}$, $4\times 10^{-3}$, $10^{-3}$, $6.4\times 10^{-4}$ and $10^{-4}$, respectively).
		\label{fig:gapwidth_vs_Kp}
		}
\end{figure}
Figure~\ref{fig:gapwidth_vs_Kp} shows $\Delta_{\rm gap}$ against the dimensionless parameter $K'$ defined by
\begin{equation}
	K'=\left( \frac{M_p}{M_{\ast}} \right)^{2} \left( \frac{h_p}{R_p} \right)^{-3} \alpha^{-1}.
	\label{eq:kp}
\end{equation}
, and there are also recoded in Table~\ref{tab:models}.
It is clear that the gap width is well scaled by the parameter $K'$.
We find an empirical formula for the gap width as
\begin{equation}
	\frac{\Delta_{\rm gap}}{R_p}= 0.41 \left( \frac{M_p}{M_{\ast}} \right)^{1/2} \left( \frac{h_p}{R_p} \right)^{-3/4} \alpha^{-1/4} = 0.41 K'^{1/4}.
	\label{eq:gapwidth}
\end{equation}

\begin{table*}
  \tbl{Our models and gap widths}{%
  \begin{tabular}{ccccc||ccccc}
\hline
$M_p/M_{\ast}$ &$h_p/R_p$ & $\alpha$ & $K'$ &  $\Delta_{\rm gap}/R_p$ & $M_p/M_{\ast}$ &$h_p/R_p$ & $\alpha$ & $K'$ &  $\Delta_{\rm gap}/R_p$ \\
\hline
$5 \times 10^{-4}$  & $1/30$ & $1 \times 10^{-2}$  & $0.68$  & $0.36$  & $5 \times 10^{-4}$  & $1/25$ & $1 \times 10^{-3}$  & $3.91$  & $0.61$  \\
$1 \times 10^{-3}$  & $1/30$ & $1 \times 10^{-2}$  & $2.71$  & $0.51$  & $1 \times 10^{-3}$  & $1/25$ & $1 \times 10^{-3}$  & $15.6$  & $0.79$  \\
$5 \times 10^{-4}$  & $1/20$ & $1 \times 10^{-2}$  & $0.20$  & $0.27$  & $3 \times 10^{-4}$  & $1/20$ & $1 \times 10^{-3}$  & $0.72$  & $0.39$  \\
$1 \times 10^{-3}$  & $1/20$ & $1 \times 10^{-2}$  & $0.80$  & $0.42$  & $5 \times 10^{-4}$  & $1/20$ & $1 \times 10^{-3}$  & $2.00$  & $0.50$  \\
$5 \times 10^{-4}$  & $1/30$ & $4 \times 10^{-3}$  & $1.69$  & $0.46$  & $7 \times 10^{-4}$  & $1/20$ & $1 \times 10^{-3}$  & $3.92$  & $0.58$  \\
$1 \times 10^{-3}$  & $1/30$ & $4 \times 10^{-3}$  & $6.77$  & $0.65$  & $1 \times 10^{-3}$  & $1/20$ & $1 \times 10^{-3}$  & $8.00$  & $0.69$  \\
$5 \times 10^{-4}$  & $1/20$ & $4 \times 10^{-3}$  & $0.50$  & $0.36$  & $1 \times 10^{-3}$  & $1/15$ & $1 \times 10^{-3}$  & $3.37$  & $0.56$  \\
$1 \times 10^{-3}$  & $1/20$ & $4 \times 10^{-3}$  & $2.00$  & $0.51$  & $2 \times 10^{-3}$  & $1/15$ & $1 \times 10^{-3}$  & $13.5$  & $0.71$  \\
$1 \times 10^{-3}$  & $1/15$ & $4 \times 10^{-3}$  & $0.84$  & $0.44$  & $1 \times 10^{-4}$  & $1/20$ & $6 \times 10^{-4}$  & $0.13$  & $0.24$  \\
$2 \times 10^{-3}$  & $1/15$ & $4 \times 10^{-3}$  & $3.37$  & $0.59$  & $5 \times 10^{-4}$  & $1/20$ & $6 \times 10^{-4}$  & $3.13$  & $0.55$  \\
$5 \times 10^{-4}$  & $1/30$ & $1 \times 10^{-3}$  & $6.77$  & $0.67$  & $1 \times 10^{-3}$  & $1/20$ & $6 \times 10^{-4}$  & $12.5$  & $0.76$  \\
$1 \times 10^{-3}$  & $1/30$ & $1 \times 10^{-3}$  & $27.1$  & $0.83$  & $1 \times 10^{-4}$  & $1/20$ & $1 \times 10^{-4}$  & $0.80$  & $0.38$  \\
$1 \times 10^{-4}$  & $1/25$ & $1 \times 10^{-3}$  & $0.16$  & $0.24$  & $5 \times 10^{-4}$  & $1/20$ & $1 \times 10^{-4}$  & $20.0$  & $0.92$  \\
\hline
\end{tabular}}\label{tab:models}
\end{table*}

As seen from Equation~(\ref{eq:gapwidth}), the gap width depends weakly on the planet mass and the viscosity, which is consistent with previous studies \citep{Varniere_Quillen_Frank2004,Duffell_MacFadyen2013} .
We also find that the disk aspect ratio affects the gap width as $h_p^{-3/4}$.
Solving Equation~(\ref{eq:gapwidth}) for $M_p/M_{\ast}$, we obtain
\begin{equation}
	\frac{M_p}{M_{\ast}}=2.1\times 10^{-3} \left( \frac{\Delta_{\rm gap}}{R_p} \right)^{2} \left( \frac{h_p}{0.05R_p} \right)^{3/2} \left( \frac{\alpha}{10^{-3}} \right)^{1/2}.
	\label{eq:mp_and_gapwidth}
\end{equation}
This equation allows us to estimate the planet mass from the observation gap width.
The planet mass strongly depends on $\Delta_{\rm gap}$ and $h_p/R_p$, as compared with $\alpha$.
Hence, if $\Delta_{\rm gap}$ and $h_p/R_p$ are measured accurately from high resolution observations, the planet mass can be well constrained.
Note that Equation~(\ref{eq:mp_and_gapwidth}) should be applied to the gap whose $\Sigma_{\min}$ is smaller than $0.45\Sigma_0$ which is the depth of the most shallow gap in Figure~\ref{fig:gapwidth_vs_Kp}.

The gap width given by Equation~(\ref{eq:gapwidth}) is reasonably consistent with that given by hydrodynamic simulations by \cite{Varniere_Quillen_Frank2004} and \cite{Duffell_MacFadyen2013}.
Their results have larger scatter, which may be partly due to the short computational time.
More detail discussions on our simulations will be described in a forthcoming paper (Kanagawa et al. in preparation).

\subsection{Test for the gap formation induced by the planet}
The mass of the planet in the gap can be estimated from the gap width (Equation~(\ref{eq:gapwidth})) or the depth (Equation~(\ref{eq:smin})).
If the width and depth give the same planet mass, it is supported that the gap is formed by a planet.
Combining Equations~(\ref{eq:smin}) and (\ref{eq:gapwidth}), we obtain the relationship among the gap width, the depth and the disk aspect ratio as 
\begin{equation}
	\frac{\Delta_{\rm gap}}{R_p} \left( \frac{\Sigma_{\min}}{\Sigma_0-\Sigma_{\min}} \right)^{1/4} \left( \frac{h_p}{R_p} \right)^{-1/2}  = 0.92.
	\label{eq:rel_width_depth}
\end{equation}
This should be satisfied for a gap created by a planet.
Note that Equation~(\ref{eq:rel_width_depth}) contains only observable quantities since the aspect ratio can be also estimated by the observed disk temperature.
When the gap width and depth are precisely observed in gas emission, Equation~(\ref{eq:rel_width_depth}) enables us to judge whether the gap is created by the planet.
For the observation of the dust thermal emission, the mass estimate from the gap width (Equation~(\ref{eq:mp_and_gapwidth})) is still useful as long as dust particles are well coupled to the gas, as discussed in the next section.

\section{Application to HL~Tau disk} \label{sec:applications}
\begin{figure}
	\begin{center}
		\resizebox{0.49\textwidth}{!}{\includegraphics{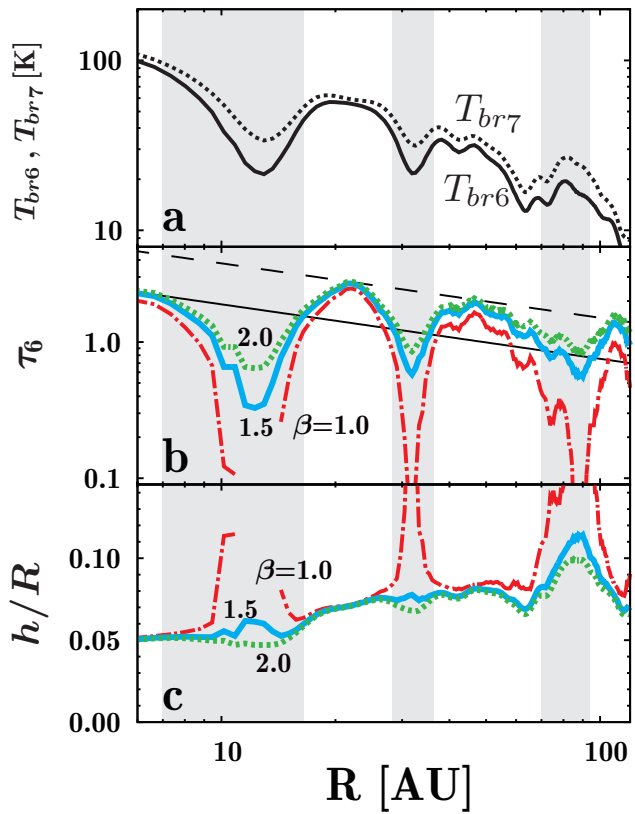}}
	\end{center}
		\caption{
		(a) Observed radial profile of the brightness temperatures of dust continuum emission in the disk of HL~Tau along the major axis. The data are averaged over $\pm 15^{\circ}$ in PA from the major axis (PA=$138^{\circ}$ and $318^{\circ}$) after the deprojection of the observed images under the assumption that the inclination angle = $46.7^{\circ}$.
		The hatched regions indicate the full width for each gap for $\beta=1.5$.
		(b) The radial profiles of the optical depth in Band~{6}.
		The thin dashed and solid lines denote the unperturbed value of the optical depth $\tau^{\rm unp}$ and $\tau^{\rm unp}/2$ for measuring the gap width (see text).
		(c) The radial profiles of the aspect ratio obtained by the disk temperature.
		\label{fig:hltau}
		}
\end{figure}
Recently, clear axisymmetric gaps in HL~Tau disk are discovered in the dust thermal emission by the Long Baseline Campaign of ALMA \citep{ALMA_HLTau2015}.
Recent hydrodynamic simulations can reproduce the observational image of HL~Tau disk by the disk--planet interaction (e.g., \cite{Dong_Zhu_Whitney2015,Dipierro_Price_Laibe_Hirsh_Cerioli_Lodato2015,Picogna_Kley2015,Jin2016}).
In Paper~I, we applied Equation~(\ref{eq:smin}) to estimate the planet mass for the HL~Tau disk.
In this study, Equation~(\ref{eq:gapwidth}) is applied to the widths of the observed gap.
As done in Paper~I, using the brightness temperatures in Band~{6} and {7}, we obtain the optical depth in Band~{6} and the gas temperature for the assumed spectral index $\beta$ (Figure~\ref{fig:hltau}).
The disk aspect ratio is calculated from the temperature by $h/R=c/(R\Omega_K)$, where $c=10^{5}(T/300\mbox{K})^{1/2}\mbox{cm/s}$.
We assume that the mass of the central star is $1M_{\odot}$.
We identify three prominent gaps in the optical depth at $R=10$AU, $30$AU and $80$AU in Figure~\ref{fig:hltau}b.
Although the gap at $80$AU can be regarded as two gaps, \cite{Dipierro_Price_Laibe_Hirsh_Cerioli_Lodato2015} pointed out that this structure can be considered as one gap with remaining dust in the horseshoe region.
Thus, we assume that the $80$AU gap is created by a single planet.

The optical depth outside the gaps can be fitted by $\tau^{\rm unp} = 9.5(R/1\mbox{AU})^{-0.4}$ in Figure~\ref{fig:hltau}b.
We regard $\tau^{\rm unp}$ as the unperturbed surface density to measure the gap width.
Note that the opacity is assumed to be constant throughout the disk.
The locations of the inner and outer gap edges, $R_{\rm in}$ and $R_{\rm out}$, are determined by intersection points with $\tau^{\rm unp}/2$ and measure the gap width as $R_{\rm out}-R_{\rm in}$.
The location of the planet, $R_p$, is simply estimated as $(R_{\rm in}+R_{\rm out})/2$.

We assume that the gap widths of the gas and dust disks are similar.
That is, it is assumed that dust particles are reasonably coupled to the disk gas and thus the dust filtration is weak.
If the dust filtration is strong, the dust surface density is enhanced at the outer edge of the gap by orders of magnitude and significantly reduced at the inner part of the disk (e.g.,\cite{Zhu2012,Dong_Zhu_Whitney2015,Picogna_Kley2015}).
In the case of relatively weak filtration, on the other hand, the gap widths of the dust disk is not altered much (see Figure~3 of \cite{Zhu2012}).
Because no significant pile--up of dust is found at the outer edge of each gap in Figure~\ref{fig:hltau}b, the assumption of the weak dust filtration would be valid for the HL~Tau disk.
\cite{Dong_Zhu_Whitney2015} also estimated the mass of the planets in the HL~Tau disk by using hydrodynamic simulations that include dust filtration for $\alpha=10^{-3}$ and $M_{\rm disk}=0.17M_{\odot}$.
Their result indeed shows that the gap width of millimeter--sized dust particles is similar (within a factor of $2$) to that of small particles tightly coupled to the gas because of relatively weak dust filtration in the massive disk of HL~Tau (see Figure~10 of \cite{Dong_Zhu_Whitney2015}), though the gap depth is much affected even in the case of the weak filtration.
Hence the gap width of the dust disk is  more suitable for estimate of the planet mass than the gap depth of the dust disk.
In addition, for particles smaller than millimeter--sized particles, gap widths (and gap depths) in gas and dust are more similar to each other. 
\cite{Jin2016} have also performed hydrodynamic simulations with 0.15 millimeter--sized particles in similar situation of \cite{Dong_Zhu_Whitney2015} ($\alpha = 10^{-3}$ and $M_{\rm disk} = 0.08 M_{\odot}$) and reproduced the observed image of HL~Tau disk.
In their simulations, the gap widths in gas and dust are very similar.

\begin{table}
  \tbl{Measured gap properties and estimated planet masses.}{%
  \begin{tabular}{cccccc}
\hline
& $R_{\rm in}$ &$R_{\rm out}$ & $\frac{\Delta_{\rm gap}}{R_p}$ & $\frac{h_p}{R_p}$ &  $M_p$ ($M_J$)  \\
&(AU)          & (AU)         &                        &           & {\footnotesize(from the width)} \\
\hline
10AU gap & $7$    & $16.5$ & $0.81$ &$0.05$ & $1.4$ \\
30AU gap & $28.5$ & $36$   & $0.23$ &$0.07$ & $0.2$ \\
80AU gap & $70$   & $94$   & $0.29$ &$0.1$  & $0.5$ \\
\hline
  \end{tabular}}\label{tab:planets}
  \begin{tabnote}
  {\footnotesize We set $\alpha=10^{-3}$, $\beta=1.5$ and $M_{\ast}=1M_{\odot}$ at the evaluation in this table.}
  \end{tabnote}
\end{table}
Table~\ref{tab:planets} shows the properties of the observed gaps and the estimated planet masses.
In this table, we set $\beta=1.5$ to obtain the optical depth and the aspect ratio,  and adopt $\alpha=10^{-3}$ for estimate of the planet masses.
The planet masses for the gaps at $10$AU, $30$AU and $80$AU are estimated to be $1.4 M_J$, $0.2 M_J$ and $0.5 M_J$ from the gap widths, respectively.
The estimated mass of the planet at $30$AU gap is consistent with that estimated from the gap depth in Paper~I.

We should note that the gap properties and the estimated masses depend on $\beta$, as seen in Figure~\ref{fig:hltau}b.
The radiative transfer model of \cite{Pinte2015} implies that $\beta\simeq 1$.
For the innermost gap, the planet mass is estimated to be $3.3M_J$ ($1.3M_J$) if $\beta=1$ ($2$) from the gap width.
For the $80$AU gap, the gap width is much more affected by the choice of $\beta$.
The disk aspect ratio can be influenced by the choice of $\beta$, which may not be neglected because the planet mass relatively strongly depends on the disk aspect ratio ($M_p \propto (h_p/R_p)^{3/2}$, see Equation~(\ref{eq:mp_and_gapwidth})).
For instance, the disk aspect ratio at the outer most gap can be changed from $0.08$ to $0.11$ if we vary $\beta$ from $2.0$ to $1.5$ (see Figure~\ref{fig:hltau}c).
In this case, the estimated mass of the planet can be changed from $0.35M_J$ to $0.57M_J$.
Therefore, accurate estimate of $\beta$ is essential in deriving the planet mass from the gap shape.
Future multi--frequency and high spatial resolution observations may constrain the planet mass better.

The location of the planet ($R_p$) can affect the mass estimate because the gap width is scaled by $R_p$ in Equation~(\ref{eq:mp_and_gapwidth}).
Although we simply estimate $R_p$ as $(R_{\rm in} + R_{\rm out})/2$ by assuming a symmetric gap, $R_p$ can be changed because the actual shape of the gap is slightly non--symmetric.
Indeed, for instance, $R_p$ for the innermost planet is set to be $\sim 13$AU in the previous simulations, which slightly larger than that in Table~\ref{tab:planets}.
If $R_p = 13$AU is adopted, the estimated mass of the planet is slightly smaller ($1.1M_J$) than that in Table~\ref{tab:planets}.

In addition to $\beta$ and $R_p$, the planet mass can also depend on the choice of the viscous parameter $\alpha$, which is highly uncertain.
However, the estimate of the planet mass varies only $\alpha^{1/2}$ (see Equation~\ref{eq:mp_and_gapwidth}) and therefore, the dependency of the planet mass on the viscous parameter is not very strong.

The relatively narrow $10$AU and $30$AU gaps are only marginally resolved with the observation.
As seen in Figure~\ref{fig:avgdens}, each gap width measured at the level of $\Sigma_0/2$ is wider than that of the bottom region, which determines the gap depth.
For the marginally resolved gap, the gap width can be accurately measured as compared with the minimum surface density of the gap.
Hence mass estimate from the gap width is less affected by the resolution.


\cite{Dong_Zhu_Whitney2015} estimated the planet masses to be $0.2M_J$ for these three gaps of the HL~Tau disk from their hydrodynamic simulations, by including dust filtration.
Their result is consistent with our estimate for the $30$AU gap.
For $10$AU gap, our estimated mass is much larger than their result.
This is probably because quantitative comparison between the model and observations is not the main focus of their work.
In their model, the gap width of millimeter--sized dust particles is $\sim 5$AU (see their Figure~10), which is about half of our measured width for the 10AU gap of the HL Tau disk.
If the gap width is half, the planet mass estimated from the width is 4 times smaller (see Equation~(\ref{eq:mp_and_gapwidth})).
This partially explains the difference between ours and theirs.
We also find that the difference of a factor of 2.5 in the planet mass estimate for the 80AU gap.
It may be due to the uncertainty in $\beta$.

Adopting 0.15 millimeter--sized particles, \cite{Jin2016} have also estimated similar masses of the planets to \cite{Dong_Zhu_Whitney2015} ($0.35M_J$, $0.17 M_J$ and $0.26 M_J$ for the innermost, middle and outermost gaps).
Their estimated masses of the planets are smaller than these given by our estimate (Table~\ref{tab:planets}), because they assumed the mass of the central star as $0.55M_{\odot}$ which is smaller than that adopted in Table~\ref{tab:planets} ($1M_{\odot}$).
Adopting $M_{\ast}=0.55M_{\odot}$, we estimate the masses of the planets as $0.77M_J, 0.11M_J$ and $0.28M_J$ for the innermost, middle and outermost gaps.
For the middle and outermost gaps, the estimated planet masses are quite similar to the result of \cite{Jin2016}.
For the innermost gap, our estimate gives the same mass of the planet as their result if the gap width is narrower in only $\sim 2$AU than that measured from Figure~\ref{fig:hltau}b.

\cite{Dipierro_Price_Laibe_Hirsh_Cerioli_Lodato2015} also derived the planet masses from a hydrodynamic simulations similar to \cite{Dong_Zhu_Whitney2015}, but by assuming a much less massive disk as $M_{\rm disk} = 0.0002M_{\odot}$.
Their result shows a strong filtration at shallow gaps for millimeter--sized particles (Figure~3 of \cite{Dipierro_Price_Laibe_Hirsh_Cerioli_Lodato2015}), in contrast to \cite{Dong_Zhu_Whitney2015} and \cite{Jin2016}.
This is reasonable because dust filtration is stronger for a less massive disk since the coupling between the gas and dust is weaker.
However, the observations suggest that the disk mass should be $\sim  0.1M_{\odot}$ for the HL~Tau disk if gas--to--dust mass ratio is $\sim 100$ \citep{Robitaille2007,ALMA_HLTau2015}.

\section{Summary} \label{sec:summary}
We have derived an empirical formula for the gap width (Equation~(\ref{eq:gapwidth})), by performing 26 runs of hydrodynamic simulation.
The gap width is expressed as a power--law function of the planet mass, the disk aspect ratio, and the viscosity.
This empirical formula enables us to estimate the planet mass from the gap width.
Paper~I presented the relationship between the gap depth and the planet mass as Equation~(\ref{eq:smin}).
If the gap is created by the planet, the masses estimated by Equations~(\ref{eq:smin}) and (\ref{eq:gapwidth}) should be consistent, and the gap width and the gap depth should satisfy Equation~(\ref{eq:rel_width_depth}).
With this, it is possible to check whether the gap is created by a planet when the gap width and depth are accurately observed in the gas emission.
For the dust thermal emission, if dust filtration is not very effective, estimate of planet mass from the gap width is still useful because the gap widths in the gas and dust disks are not so different.

We have applied the empirical formula for the gap width to the gaps in the HL~Tau disk observed in dust thermal emission by ALMA.
We have estimated the mass of planets in the gaps at $10$AU, $30$AU and $80$AU as $1.4 M_J$, $0.2 M_J$ and $0.5M_J$, respectively, assuming $M_{\ast}=1M_{\odot}$.
For the innermost gap, the whole structure may not be completely resolved by the observation and measuring the gap depth is difficult.
The dust filtration alters the gap depth more than the gap width.
The estimate form the gap width gives us a more accurate planet mass than that from the gap depth.
Our estimate depends on the particle size of dusts (i.e., the dust opacity index of $\beta$) and the disk model for the dust filtration.
More sophisticated models in the HL~Tau disk would improve the above estimates of the planet mass.
If the gap is observed in gas emission, we can constrain a planet mass from the gap depth and width, without uncertainty of the dust and disk models.

\begin{ack}
We thank Ruobing Dong for giving us his data.
This paper makes use of the following ALMA data: ADS/JAO.ALMA\#2011.0.00015.SV.  
ALMA is a partnership of ESO (representing its member states), NSF (USA)
and NINS (Japan), together with NRC (Canada) and NSC and ASIAA (Taiwan),  
in cooperation with the Republic of Chile. The Joint ALMA Observatory is 
operated by ESO, AUI/NRAO and NAOJ.
This work was supported by JSPS KAKENHI Grant Numbers 23103004, 26103701, 26800106 and Polish National Science Centre MAESTRO grant DEC- 2012/06/A/ST9/00276.
KDK was supported by the ALMA Japan Research Grant of NAOJ Chile Observatory, NAOJ-ALMA-0135.
Numerical computations were carried out on Cray XC30 at Center for Computational Astrophysics, National Astronomical Observatory of Japan and the Pan-Okhotsk Information System at the Institute of Low Temperature Science, Hokkaido University.
\end{ack}


\end{document}